\documentclass[aps,preprint]{revtex4-1}
\usepackage{graphicx}
\usepackage{amsmath}
\usepackage{makeidx}
\usepackage{amsfonts}
\usepackage{amssymb}

\begin{document}

\title{Characterizing the parent Hamiltonians for a complete set of orthogonal wave functions: An inverse quantum problem}

\author{A. Ramezanpour}
\email{aramezanpour@gmail.com}
\affiliation{Department of Physics, University of Neyshabur, Neyshabur, Iran}
\date{\today}

\date{\today}

\begin{abstract}
We study the inverse problem of constructing an appropriate Hamiltonian from a physically reasonable set of orthogonal wave functions for a quantum spin system. Usually, we are given a local Hamiltonian and our goal is to characterize the relevant wave functions and energies (the spectrum) of the system. Here, we take the opposite approach; starting from a reasonable collection of orthogonal wave functions, we try to characterize the associated parent Hamiltonians, to see how the wave functions and the energy values affect the structure of the parent Hamiltonian. Specifically, we obtain (quasi) local Hamiltonians by a complete set of (multilayer) product states and a local mapping of the energy values to the wave functions. On the other hand, a complete set of tree wave functions (having a tree structure) results to nonlocal Hamiltonians and operators which flip simultaneously all the spins in a single branch of the tree graph. We observe that even for a given set of basis states, the energy spectrum can significantly change the nature of interactions in the Hamiltonian. These effects can be exploited in a quantum engineering problem optimizing an objective functional of the Hamiltonian.
\end{abstract}


\maketitle

\section{Introduction}\label{S0}
The Hamiltonian of an interacting system governs all the physical behaviors of the system; from the quantum and thermal expectation of the physical observables to the time evolution of the state of system. In a direct physical problem, the Hamiltonian is given with the aim of inferring the relevant expectation values either directly by diagonalizing the Hamiltonian or indirectly, e.g., by a Monte Carlo simulation algorithm. In an inverse problem, we are given some physical expectations and seek an appropriate Hamiltonian, which reproduces the set of desirable features. The Hamiltonian can be completely specified by the whole set of its eigenstates (wave functions) and eigenvalues (the energy spectrum or function). A succinct representation of the wave and energy functions then could be very helpful in exploring and optimizing in the exponentially large space of the possible Hamiltonians (e.g., in classification of quantum phases \cite{CGW-prb-2011,SPC-prb-2011}). An advantage of this approach is that we can directly control the spectrum (e.g., the energy gap), and we can choose to work with computationally tractable wave functions. 

In this paper, we take a physically reasonable ensemble of orthogonal wave functions and study the effects of the wave and energy functions on the structure (e.g., locality) of the parent Hamiltonians. We know that the ground state of a fully-connected (mean-field) quantum Ising model with uniform (homogeneous) couplings is a product state in the thermodynamic limit \cite{R-nphys-2007,KLC-pra-2013,SNCS-japan-2000}. It would then be interesting to characterize the set of parent Hamiltonians which we can obtain by a complete set of orthogonal product states or multilayer wave functions of product states \cite{R-epl-2013}. Here, we show that an appropriate (local) mapping of the energy values to the wave functions leads to (quasi) local parent Hamiltonians. This provides another example for construction and study of the fully many-body-localized systems \cite{SPA-prl-2013,HNO-prb-2014,PC-arxiv-2014,CCKAV-arxiv-2014,RMS-np-2015,PKCS-arxiv-2015}.        

An important class of wave functions are obtained by including two-body interactions in the wave function.
This includes the famous Jastrow wave functions \cite{J-pr-1955} and the matrix product states \cite{FNW-cmp-1992,VMC-advp-2008}, which have been successfully employed to describe the quantum state of many interacting physical systems \cite{DM-prb-1980,L-prl-1983,HE-prl-1988,CKUC-prb-2009,VC-prb-2006,PVCW-qinf-2008}. Tree wave functions, with interaction graphs that have a tree structure, provide another ensemble of computationally tractable states, which are very close to the Jastrow wave functions.  The parent Hamiltonians that we obtain by such a complete set of orthogonal wave functions are nonlocal, with global operators that act simultaneously on all spins in a single branch of the tree graph. 
 
Here, we shall restrict ourselves to spin systems, but the approach is also applicable to other quantum systems. One, of course, needs a complete set of orthogonal wave functions (basis states) for the quantum system of variables. Then, any real energy function of the states defines a Hamiltonian for the system. The problem is to find out how the basis states and the energy values affect the structure of the Hamiltonian. It is obviously better to work with computationally tractable wave functions which allow us to compute easily the physical properties of the system. That is why we consider the simple product and tree states for the spin systems.         

In this study, we are indeed characterizing the ensemble of the parent Hamiltonians
for a complete set of simple wave functions. Given the basis states, the associated energy
values are considered as free variables which can be adjusted to construct different Hamiltonians.          
For given functional form of the basis states and the energy values, the free parameters in the states and the energy function can still be varied to control the strength of the interactions in the parent Hamiltonian.     
 
The paper is organized as follows. Section \ref{S1} gives the main definitions. In Sec. \ref{S2}, we characterize the parent Hamiltonians we obtain by the (one-layer) product states. In Sec. \ref{S3}, we consider the tree states with symmetric two-body interactions in the wave functions. In Sec. \ref{S4}, we briefly describe the construction of multilayer wave functions, and study specifically the structure of the parent Hamiltonians for the two-layer product states. Section \ref{S5} gives the concluding remarks.

\section{Main definitions}\label{S1}
We consider a system of quantum spins $\{\sigma_i^{x,y,z}|i=1,\dots,N\}$, where $\sigma_i^{x,y,z}$ are the known Pauli matrices.  We will represent the wave functions of the system by $|\psi \rangle=\sum_{\boldsymbol\sigma} \psi(\boldsymbol\sigma)|\boldsymbol\sigma \rangle$ using the standard computational basis $|\boldsymbol\sigma \rangle$. Here $\boldsymbol\sigma \equiv \{\sigma_1,\dots,\sigma_N\}$ shows the $\sigma^z$ values of the spins $i=1,\dots,N$.
    
We start from an arbitrary wave function $|\mathbf{0} \rangle=\sum_{\boldsymbol\sigma} \psi(\mathbf{0};\boldsymbol\sigma)|\boldsymbol\sigma \rangle$ with a set of parameters denoted by $\mathbf{P}(\mathbf{0})$. Suppose we have constructed a set of orthonormal wave functions $|\mathbf{s} \rangle$ with their own parameters $\mathbf{P}(\mathbf{s})$, where $\langle \mathbf{s}|\mathbf{0} \rangle=0$ for $\mathbf{s}\neq \mathbf{0}$, and $\langle \mathbf{s}' |\mathbf{s} \rangle=\delta_{\mathbf{s},\mathbf{s}'}$. It is obvious that $|\mathbf{0} \rangle$ is the ground state of a (globally) frustration-free Hamiltonian \cite{BDOT-qinf-2008,FSWCP-cmath-2015},  
\begin{align}
H_P(\mathbf{0}) \equiv \sum_{\mathbf{s} \neq \mathbf{0}} \lambda(\mathbf{s}) |\mathbf{s} \rangle\langle \mathbf{s}|,
\end{align}
for real and nonnegative coefficients $\lambda(\mathbf{s})$. Furthermore, state $|\mathbf{s} \rangle$ is an eigenstate of $H_p(\mathbf{0})$ with eigenvalue $\lambda(\mathbf{s})$. For a complete set of states $\sum_{\mathbf{s}} |\mathbf{s} \rangle\langle \mathbf{s}| =1$, we are sure that these states can be used to represent the whole spectrum of the Hamiltonian. Then, for real energy functions $\lambda(\mathbf{s})$, we have the parent Hamiltonian
\begin{align}
H_P \equiv  \sum_{\mathbf{s}} \lambda(\mathbf{s}) |\mathbf{s} \rangle\langle \mathbf{s}|,
\end{align}
with all the eigenstates $|\mathbf{s} \rangle$ and eigenvalues $\lambda(\mathbf{s})$. Here the ground state is the one minimizing  $\lambda(\mathbf{s})$. We stress that here the parent Hamiltonian is completely defined if we have both
the eigenstates and eigenvalues. Therefore, even after fixing the basis states, we need to know the energy values to completely specify the parent Hamiltonian.

Notice that the above Hamiltonian could be nonlocal in the standard representation; suppose $|\mathbf{s} \rangle=\sum_{\boldsymbol\sigma} \psi(\mathbf{s};\boldsymbol\sigma)|\boldsymbol\sigma \rangle$, then the matrix elements of the parent Hamiltonian are
\begin{align}
\langle \boldsymbol\sigma |H_P| \boldsymbol\sigma'\rangle=\sum_{\mathbf{s}} \lambda(\mathbf{s}) \psi(\mathbf{s};\boldsymbol\sigma)\psi^*(\mathbf{s};\boldsymbol\sigma')\equiv H_P(\boldsymbol\sigma,\boldsymbol\sigma').
\end{align}
In a $k$-local Hamiltonian, there could be spin interactions between at most $k$ spins ($k$-body interactions). For such a Hamiltonian, $H_P(\boldsymbol\sigma,\boldsymbol\sigma')=0$ when the number of different spins in the two spin configurations (the Hamming distance) is greater than $k$. The Hamiltonian is called quasilocal if the strength of $k$-body interactions decreases exponentially with $k$. The question then is how the wave functions $\psi(\mathbf{s};\boldsymbol\sigma)$ and energy function $\lambda(\mathbf{s})$ affect the locality and other features of the parent Hamiltonian.

\section{Product wave functions}\label{S2}
In this section, we start from a complete set of product states for a quantum spin system.
Then, we show that an appropriate mapping of the energy values to these states results to 
a local parent Hamiltonian for the system. The analysis will provide insight for constructing a computationally tractable
Hamiltonian with adjustable (few-body) interactions.

Consider the simple product states
\begin{align}
\psi(\mathbf{0};\boldsymbol\sigma)= e^{\hat{i}\Theta(\boldsymbol\sigma)}\prod_i \left(\frac{e^{B_i\sigma_i/2}}{\sqrt{2\cosh B_i^R}} \right),
\end{align}
where the parameters $B_i=B_i^R+\hat{i}B_i^I$ are in general complex numbers. We have also included an arbitrary but real phase $\Theta(\boldsymbol\sigma)$ to the wave function; to be specific, we work with a local phase $\Theta(\boldsymbol\sigma) \equiv \sum_{i<j} \Lambda_{ij}\sigma_i\sigma_j/2$, where the sum is over all the pairs of spins. We obtain a set of orthonormal product states $|\mathbf{s} \rangle \equiv |s_1,\dots,s_N \rangle$, with occupation numbers $s_i \in \{0,1\}$ that show the absence ($s_i=0$) or presence ($s_i=1$) of a local "excitation" at site $i$  \cite{BR-jstat-2013,R-epl-2013}. Let us represent the above states by $|\mathbf{s} \rangle=\sum_{\boldsymbol\sigma} \psi(\mathbf{s};\boldsymbol\sigma)|\boldsymbol\sigma \rangle$. The dependence on $\mathbf{s}$ enters only in the parameters $B_i(s_i)=(1-2s_i)B_i^R+\hat{i}(B_i^I+s_i\pi)$, that is,
\begin{align}
\psi(\mathbf{s};\boldsymbol\sigma) = e^{\hat{i}\Theta(\boldsymbol\sigma)}\prod_i \left(\frac{e^{B_i(s_i)\sigma_i/2}}{\sqrt{2\cosh B_i^R }} \right).
\end{align}
Then, one can easily prove the following orthogonality relations:
\begin{align}
\sum_{\boldsymbol\sigma} \psi^*(\mathbf{s};\boldsymbol\sigma)\psi(\mathbf{s}';\boldsymbol\sigma) &=\delta_{\mathbf{s},\mathbf{s}'},\\
\sum_{\mathbf{s}} \psi^*(\mathbf{s};\boldsymbol\sigma)\psi(\mathbf{s};\boldsymbol\sigma') &=\delta_{\boldsymbol\sigma,\boldsymbol\sigma'}.
\end{align}
In fact the $2^N$ locally excited states are mutually orthonormal and $\sum_{\mathbf{s}} |\mathbf{s} \rangle\langle \mathbf{s}|=1$.
Note that here the parameters $\mathbf{P}(\mathbf{0})=\{B_i,\Lambda_{ij}|i=1,\dots,N, j<i\}$ are enough to identify the whole set of parameters $\mathbf{P}(\mathbf{s})$.

\subsection{Characterizing the parent Hamiltonian}
The matrix elements of the parent Hamiltonian are given by 
\begin{multline}
H_P(\boldsymbol\sigma,\boldsymbol\sigma') = \sum_{\mathbf{s}} \lambda(\mathbf{s}) \psi(\mathbf{s};\boldsymbol\sigma)\psi^*(\mathbf{s};\boldsymbol\sigma') \\
= e^{\hat{i}[\Theta(\boldsymbol\sigma)-\Theta(\boldsymbol\sigma')]}\times  \left( \frac{1}{Z}\sum_{\mathbf{s}} \lambda(\mathbf{s}) e^{\sum_i [B_i(s_i)\sigma_i+B_i^*(s_i)\sigma_i']/2}\right)
\equiv e^{\hat{i}\Phi(\boldsymbol\sigma,\boldsymbol\sigma')}\mathsf{H}(\boldsymbol\tau),
\end{multline}
where $Z\equiv \prod_i \left(2\cosh B_i^R \right)$.
Here, we defined the phase and the transformed function
\begin{align}
\Phi(\boldsymbol\sigma,\boldsymbol\sigma') &\equiv \Theta(\boldsymbol\sigma)-\Theta(\boldsymbol\sigma')+\sum_i(B_i^I+\pi/2)(\frac{\sigma_i-\sigma_i'}{2}),\\
\mathsf{H}(\boldsymbol\tau) &\equiv \frac{1}{Z}\sum_{\mathbf{s}} \lambda(\mathbf{s}) e^{\sum_i \tau_i(1-2s_i)},
\end{align}
with $\tau_i \equiv B_i^R(\sigma_i+\sigma_i')/2-\hat{i}\pi(\sigma_i-\sigma_i')/4$.
 
For example, let us consider the local energy function 
\begin{align}
\lambda(\mathbf{s})=-\sum_i h_i(1-2s_i)-\sum_{i<j}J_{ij}(1-2s_i)(1-2s_j).
\end{align}
Then, for the transformed energy function, we obtain
\begin{multline}
\mathsf{H}(\boldsymbol\tau)= \frac{1}{\prod_i \cosh B_i^R}\Big\{-\sum_i h_i\sinh \tau_i \prod_{j\neq i}\cosh \tau_j -\sum_{i<j}J_{ij}\sinh \tau_i\sinh \tau_j \prod_{k\neq i,j}\cosh \tau_k \Big\}.
\end{multline}
Note that $\cosh \tau_i =0$ for $\sigma_i' \neq \sigma_i$. Therefore, the first terms in the above expression are nonzero only if $\boldsymbol\sigma'$ differs from $\boldsymbol\sigma$ at most in one spin; a configuration that is different from $\boldsymbol\sigma$ only at site $i$ is shown by $\boldsymbol\sigma'=\boldsymbol\sigma^{-i}$. Similarly, the second terms are nonzero only if the two spin configurations are different at most in two spins; we use $\boldsymbol\sigma'=\boldsymbol\sigma^{-ij}$ for the spin configuration that is different from $\boldsymbol\sigma$ only at sites $i$ and $j$.

In this way, for the matrix elements of the parent Hamiltonian, we find
\begin{align}
H_P(\boldsymbol\sigma,\boldsymbol\sigma) &= -\sum_i (h_i \tanh B_i^R) \sigma_i- \sum_{i<j} (J_{ij} \tanh B_i^R \tanh B_j^R) \sigma_i\sigma_j,\\
H_P(\boldsymbol\sigma,\boldsymbol\sigma^{-i}) &= e^{\hat{i}[\sum_{j\neq i}\Lambda_{ij}\sigma_i\sigma_j+B_i^I\sigma_i]}\frac{1}{\cosh B_i^R}\left(-h_i- \sum_{j\neq i} (J_{ij} \tanh B_j^R) \sigma_j \right),\\
H_P(\boldsymbol\sigma,\boldsymbol\sigma^{-ij}) &= e^{\hat{i}[\sum_{k\neq i,j}(\Lambda_{ik}\sigma_i+\Lambda_{jk}\sigma_j)\sigma_k+B_i^I\sigma_i+B_j^I\sigma_j]}\frac{-J_{ij}}{\cosh B_i^R\cosh B_j^R}.
\end{align}
And, $H_P(\boldsymbol\sigma,\boldsymbol\sigma')=0$ if $\boldsymbol\sigma$ and $\boldsymbol\sigma'$ are different in more than two spins. This means that we can write the Hamiltonian as follows:
\begin{multline}
H_P= \left(-\sum_i h_i^z \sigma_i^z- \sum_{i<j} J_{ij}^{zz} \sigma_i^z\sigma_j^z \right)\\
+\sum_i e^{\hat{i}[\sum_{j\neq i}\Lambda_{ij}\sigma_i^z\sigma_j^z+B_i^I\sigma_i^z]}\left(-h_i^x\sigma_i^x - \sum_{j\neq i} J_{ij}^{xz} \sigma_i^x\sigma_j^z \right) \\
- \sum_{i<j} e^{\hat{i}[\sum_{k\neq i,j}(\Lambda_{ik}\sigma_i^z+\Lambda_{jk}\sigma_j^z)\sigma_k^z+B_i^I\sigma_i^z+B_j^I\sigma_j^z]}J_{ij}^{xx} \sigma_i^x\sigma_j^x,
\end{multline}
where
\begin{align}
h_i^z &\equiv h_i \tanh B_i^R,\hskip1cm J_{ij}^{zz} \equiv J_{ij} \tanh B_i^R \tanh B_j^R,\\
h_i^x &\equiv \frac{h_i}{\cosh B_i^R},\hskip0.5cm J_{ij}^{xz} \equiv \frac{J_{ij}\tanh B_j^R}{\cosh B_i^R},\hskip0.5cm 
J_{ij}^{xx} \equiv \frac{J_{ij}}{\cosh B_i^R\cosh B_j^R}.
\end{align}

Now, it is easy to write the parent Hamiltonian for the following energy function,
\begin{align}
\lambda(\mathbf{s})=- \sum_{i_1<i_2<\dots<i_k}J_{i_1i_2\dots i_k}\prod_{l=1}^k(1-2s_{i_l}).
\end{align}
In this case, the elements $H_P(\boldsymbol\sigma,\boldsymbol\sigma')$ are zero if the Hamming distance of the two spin configurations is greater than $k$. In general, we can look for an optimal spectrum $\lambda(\mathbf{s})$ minimizing the distance 
\begin{align}
\sum_{\boldsymbol\tau}\left(\frac{1}{Z}\sum_{\mathbf{s}} \lambda(\mathbf{s}) e^{\sum_i \tau_i(1-2s_i)}-\mathsf{H}_0(\boldsymbol\tau) \right)^2,
\end{align}
of the transformed energy function from a desirable function $\mathsf{H}_0(\boldsymbol\tau)$.

\section{Symmetric tree wave functions}\label{S3}
As another example, we consider the symmetric tree states defined on a tree graph $\mathcal{T}$,
\begin{align}
\psi(\mathbf{0};\boldsymbol\sigma)= \frac{e^{\hat{i}\Theta(\boldsymbol\sigma)}}{\sqrt{2}}\prod_{(ij) \in \mathcal{T}} \left(\frac{e^{K_{ij}\sigma_i\sigma_j/2}}{\sqrt{2\cosh K_{ij}^R}} \right).
\end{align}
The phase $\Theta(\boldsymbol\sigma)$ is arbitrary but a real number; to be specific we take $\Theta(\boldsymbol\sigma)=\sum_i \Lambda_i \sigma_i/2$. The $\Lambda_i$ and the complex couplings $K_{ij}=K_{ij}^R+\hat{i}K_{ij}^I$ define the parameters $\mathbf{P}(\mathbf{0})$.  
From the above state, we obtain a set of orthonormal tree states $|\mathbf{s} \rangle$, with $s_{ij} \in \{0,1\}$ to show the absence or presence of a local "excitation" at edge $(ij)$ \cite{BR-jstat-2013,R-epl-2013}. The dependence on $\mathbf{s}$ enters only in the parameters $K_{ij}(s_{ij})=(1-2s_{ij})K_{ij}^R+\hat{i}(K_{ij}^I+s_{ij}\pi)$, that is,
\begin{align}
\psi(\mathbf{s};\boldsymbol\sigma)= \frac{e^{\hat{i}\Theta(\boldsymbol\sigma)}}{\sqrt{2}}\prod_{(ij) \in \mathcal{T}} \left(\frac{e^{K_{ij}(s_{ij})\sigma_i\sigma_j/2}}{\sqrt{2\cosh K_{ij}^R}} \right).
\end{align}
Here, we have only $2^{N-1}$ of such locally excited states, each one being a symmetric state spanning both the positive and negative sectors of the configuration space.   

Let us use $|\mathbf{s};\pm \rangle$ for the contribution of the positive and negative sectors of the configuration space to the wave functions, i.e.,
\begin{align}
|\mathbf{s} \rangle=\frac{1}{\sqrt{2}}\left(|\mathbf{s};+ \rangle +|\mathbf{s};- \rangle \right).
\end{align}
It is obvious that $\langle \mathbf{s};+|\mathbf{s}';- \rangle=0$, and by symmetry $\langle \mathbf{s};+|\mathbf{s};+ \rangle=\langle \mathbf{s};-|\mathbf{s};- \rangle$ for any $\mathbf{s},\mathbf{s}'$. Moreover, since $\langle \mathbf{s}|\mathbf{s}'\rangle=\delta_{\mathbf{s},\mathbf{s}'}$, we have
\begin{align}
\langle \mathbf{s};+|\mathbf{s}';+ \rangle+\langle \mathbf{s};-|\mathbf{s}';- \rangle=2\delta_{\mathbf{s},\mathbf{s}'}.
\end{align}
Then, for each state $|\mathbf{s}\rangle$, we define the two orthogonal states,
\begin{align}
|\mathbf{s};\theta=\pm \pi/2 \rangle=\frac{1}{\sqrt{2}}\left(|\mathbf{s};+ \rangle + e^{\hat{i}\theta}|\mathbf{s};- \rangle \right).
\end{align}
From the above equations, we find the following orthogonality relations: 
\begin{align}
\langle \mathbf{s}; \theta|\mathbf{s}';\theta'\rangle=\delta_{\mathbf{s},\mathbf{s}'}\delta_{\theta,\theta'}.
\end{align}
In this way, we obtain a complete set of $2^N$ symmetric tree states including the initial one $|\mathbf{0}\rangle$. Now, the parent Hamiltonian reads as
\begin{align}
H_P=\sum_{\theta=\pm \pi/2} \sum_{\mathbf{s}} \lambda(\mathbf{s}) |\mathbf{s};\theta \rangle\langle \mathbf{s};\theta|,
\end{align}
with the two ground states $|\mathbf{s}^*; \theta=\pm \pi/2 \rangle$ minimizing $\lambda(\mathbf{s})$.

\subsection{Characterizing the parent Hamiltonian}
First note that
\begin{align}
\langle \boldsymbol\sigma |\mathbf{s};\theta \rangle = \psi(\mathbf{s};\boldsymbol\sigma) (\delta_{+}+e^{\hat{i}\theta}\delta_{-})\equiv \psi_{\theta}(\mathbf{s};\boldsymbol\sigma),
\end{align}
where $\delta_{+}$ is one if magnetization $\sum_i \sigma_i > 0$; otherwise, it is zero. Similarly $\delta_{-}$ is one only if $\sum_i \sigma_i < 0$; we assume that $N$ is an odd number. Then, the matrix elements of the Hamiltonian are given by 
\begin{align}
H_P(\boldsymbol\sigma,\boldsymbol\sigma') = \sum_{\theta=\pm \pi/2}\sum_{\mathbf{s}} \lambda(\mathbf{s}) \psi_{\theta}(\mathbf{s};\boldsymbol\sigma)\psi_{\theta}^*(\mathbf{s};\boldsymbol\sigma')
= e^{\hat{i}\Phi(\boldsymbol\sigma,\boldsymbol\sigma')} \mathsf{H}(\boldsymbol\tau)(\delta_+\delta_+'+\delta_-\delta_-'),
\end{align}
where $Z\equiv \prod_{(ij) \in \mathcal{T}} \left(2\cosh K_{ij}^R \right)$.
Here, we defined the phase and the transformed energy function,
\begin{align}
\Phi(\boldsymbol\sigma,\boldsymbol\sigma') &\equiv \Theta(\boldsymbol\sigma)-\Theta(\boldsymbol\sigma')+\sum_{(ij) \in \mathcal{T}}(K_{ij}^I+\pi/2)(\frac{\sigma_{ij}-\sigma_{ij}'}{2}),\\
\mathsf{H}(\boldsymbol\tau) &\equiv \frac{1}{Z}\sum_{\mathbf{s}} \lambda(\mathbf{s}) e^{\sum_{(ij) \in \mathcal{T}} \tau_{ij}(1-2s_{ij})},
\end{align} 
with $\tau_{ij} \equiv K_{ij}^R(\sigma_{ij}+\sigma_{ij}')/2-\hat{i}\pi(\sigma_{ij}-\sigma_{ij}')/4$. 
To shorten the notation, we also defined $\sigma_{ij}\equiv \sigma_i\sigma_j$. Moreover, the $\delta_{\pm}'$ are defined with respect to the prime spin configuration $\boldsymbol\sigma'$. Note that the matrix elements connecting two symmetry-related configurations are zero.

As an example, consider the simple energy function 
\begin{align}
\lambda(\mathbf{s})=-\sum_{(ij) \in \mathcal{T}} h_{ij}(1-2s_{ij}).
\end{align}
Then, using the definition of the transformed energy function, we find
\begin{align}
\mathsf{H}(\boldsymbol\tau)= \frac{1}{\prod_{(ij) \in \mathcal{T}} \cosh K_{ij}^R}\left(-\sum_{(ij) \in \mathcal{T}} h_{ij} \sinh \tau_{ij} \prod_{(kl)\neq (ij) \in \mathcal{T}}\cosh \tau_{kl} \right).
\end{align}
We note that $\cosh \tau_{ij}=0$ for $\sigma_{ij}' \neq \sigma_{ij}$. Consequently, with the above energy function, $\mathsf{H}(\boldsymbol\tau)$ is nonzero only if the two spin configurations differ at most in one link, say $\sigma_{ij}'=-\sigma_{ij}$ for some $(ij) \in \mathcal{T}$. The resulting parent Hamiltonian, which is restricted to the positive or negative sector of the Hilbert space, reads as
\begin{multline}
H_P= -\sum_{(ij) \in \mathcal{T}} J_{ij}^{zz} \sigma_i^z\sigma_j^z \\
-\sum_{(ij) \in \mathcal{T}}e^{\hat{i}K_{ij}^I\sigma_i^z\sigma_j^z} \left(e^{\hat{i}\sum_{k \in \mathsf{T}_{i\to j}} \Lambda_k \sigma_k^z}h_{i\to j}^x \sigma_{i\to j}^x+e^{\hat{i}\sum_{k \in \mathsf{T}_{j\to i}} \Lambda_k \sigma_k^z}h_{j\to i}^x \sigma_{j\to i}^x \right),
\end{multline}
where  
\begin{align}
J_{ij}^{zz} &\equiv h_{ij}\tanh K_{ij}^R,\hskip1cm h_{i\to j}^x =h_{j\to i}^x \equiv \frac{h_{ij}}{\cosh K_{ij}^R}.
\end{align}
Here, we defined the global operators $\sigma_{i\to j}^x \equiv \prod_{k \in \mathcal{T}_{i\to j}} \sigma_k^x$, which flips simultaneously all the spins in the cavity tree $\mathcal{T}_{i\to j}$ defined recursively by $\mathcal{T}_{i\to j}=i \cup_{k \in \partial i\setminus j} \mathcal{T}_{k\to i}$. Here $\partial i$ denotes the set of neighbors of site $i$ in the tree.

In the same way, one can obtain the parent Hamiltonian for an energy function like
\begin{align}
\lambda(\mathbf{s})=-\sum_{i}\sum_{k<l : k,l \in \partial i} J_{ikl}(1-2s_{ik})(1-2s_{il}),
\end{align}
where the transformed energy function is given by 
\begin{align}
\mathsf{H}(\boldsymbol\tau)= \frac{1}{\prod_{(ij) \in \mathcal{T}} \cosh K_{ij}^R}\left(-\sum_{i}\sum_{k<l : k,l \in \partial i} J_{ikl}\sinh \tau_{ik}\sinh \tau_{il} \prod_{(mn)\neq \{(ik),(il)\} \in \mathcal{T}}\cosh \tau_{mn} \right).
\end{align}

\section{Multilayer wave functions of product states}\label{S4}
In this section, we study the parent Hamiltonians obtained by an orthonormal set of multilayer wave functions \cite{R-epl-2013}.
A multilayer wave function is constructed by a coupling of simpler wave functions in two or more layers.   
In the following, we consider multilayer wave functions of product states, which are more structured than the (one-layer) product states and still computationally tractable for small numbers of the layers. In Ref. \cite{R-epl-2013}, we showed that for such a $t$-layer wave function, the two-spin correlations $\langle \sigma_i^z \sigma_j^z\rangle - \langle \sigma_i^z\rangle \langle \sigma_j^z \rangle$ could be nonzero if the two spins have a distance $d_{ij}< 2t$.       

More precisely, we define a $(t+1)$-layer wave function $|\boldsymbol\sigma_t\rangle$ recursively by
\begin{align}
|\boldsymbol\sigma_t\rangle = \sum_{\boldsymbol\sigma_{t-1}}\psi_{t-1}(\boldsymbol\sigma_{t};\boldsymbol\sigma_{t-1})|\boldsymbol\sigma_{t-1}\rangle,
\end{align}
where the $|\boldsymbol\sigma_0\rangle$ are the physical spin states in the standard representation. Also, the wave functions in each step are orthonormal $\langle \boldsymbol\sigma_t|\boldsymbol\sigma_t'\rangle=\delta_{\boldsymbol\sigma_t,\boldsymbol\sigma_t'}$. 
In each layer, we start from a product wave function, 
\begin{align}
\psi_l(+;\boldsymbol\sigma_{l})=e^{\hat{i}\Theta^l(\boldsymbol\sigma_{l})}\prod_i \frac{e^{B_i^l\sigma_{l,i}/2}}{\sqrt{2\cosh B_i^{l,R}}},
\end{align}
with complex fields $B_i^l=B_i^{l,R}+\hat{i}B_i^{l,I}$ and phase $\Theta^l(\boldsymbol\sigma_{l})=\sum_{i<j} \Lambda_{ij}^l\sigma_{l,i}\sigma_{l,j}/2$, 
for some real couplings $\Lambda_{ij}^l$. Then, we write the locally "excited" states as
\begin{align}
\psi_l(\boldsymbol\sigma_{l+1};\boldsymbol\sigma_{l})=e^{\hat{i}\Theta^l(\boldsymbol\sigma_{l})}\prod_i \frac{e^{B_i^l(\sigma_{l+1,i})\sigma_{l,i}/2}}{\sqrt{2\cosh B_i^{l,R}}},
\end{align}
where $B_i^l(\sigma=\pm 1)=\sigma B_i^{l,R}+\hat{i}(B_i^{l,I}+(1-\sigma)\pi/2)$, and the phase $\Theta^l(\boldsymbol\sigma_{l})$ remains the same for all the $\boldsymbol\sigma_{l+1}$ configurations. Here $\boldsymbol\sigma_{l+1}$ determines the configuration of local excitations; $\sigma_{l+1,i}=-1(+1)$ shows the presence (absence) of a local excitation at site $i$. Notice that in each layer, we need only the parameters $\mathbf{P}_l(+)=\{B_i^l,\Lambda_{ij}^l|i=1,\dots,N, j<i\}$ to identify all the parameters $\mathbf{P}_l(\boldsymbol\sigma_{l+1})$ in that layer.

The intermediate wave functions $\psi_l(\boldsymbol\sigma_{l+1};\boldsymbol\sigma_{l})$ satisfy the following relations: 
\begin{align}
\sum_{\boldsymbol\sigma_{l}}\psi_{l}^*(\boldsymbol\sigma_{l+1};\boldsymbol\sigma_{l})\psi_{l}(\boldsymbol\sigma_{l+1}';\boldsymbol\sigma_{l}) &=\delta_{\boldsymbol\sigma_{l+1},\boldsymbol\sigma_{l+1}'},\\
\sum_{\boldsymbol\sigma_{l+1}}\psi_{l}^*(\boldsymbol\sigma_{l+1};\boldsymbol\sigma_{l})\psi_{l}(\boldsymbol\sigma_{l+1};\boldsymbol\sigma_{l}') &=\delta_{\boldsymbol\sigma_{l},\boldsymbol\sigma_{l}'}.
\end{align}
The first relation comes from the orthogonality of the local excitations, and the second ensures that the new states make a complete basis. Both the relations can be checked directly for the above product wave functions.

For a given number of layers, the parent Hamiltonian $H_P(t)$ is
\begin{align}
H_P(t)=\sum_{\boldsymbol\sigma_t} \lambda(\boldsymbol\sigma_t) |\boldsymbol\sigma_t\rangle\langle \boldsymbol\sigma_t|.
\end{align}
This parent Hamiltonian has eigenstates $|\boldsymbol\sigma_t\rangle$ of energies $\lambda(\boldsymbol\sigma_t)$.
Here, the matrix elements of the Hamiltonian are given by
\begin{align}
H_P(\boldsymbol\sigma_0,\boldsymbol\sigma_0')=\sum_{\boldsymbol\sigma_t} \lambda(\boldsymbol\sigma_t) \psi_t(\boldsymbol\sigma_t;\boldsymbol\sigma_0)\psi_t^*(\boldsymbol\sigma_t;\boldsymbol\sigma_0'),
\end{align}
with the multilayer wave function 
\begin{align}
\psi_t(\boldsymbol\sigma_t;\boldsymbol\sigma_0)=\sum_{\boldsymbol\sigma_1,\dots,\boldsymbol\sigma_{t-1}} \psi_{t-1}(\boldsymbol\sigma_t;\boldsymbol\sigma_{t-1}) \cdots \psi_{0}(\boldsymbol\sigma_1;\boldsymbol\sigma_0).
\end{align}

\subsection{Characterizing the parent Hamiltonian}
Let us, for simplicity, consider the two-layer wave functions, to see how the Hamiltonian changes with the energy values.  
From the above equations for the parent Hamiltonian and the two-layer wave functions, we have 
\begin{align}
H_P(\boldsymbol\sigma_0,\boldsymbol\sigma_0')=\sum_{\boldsymbol\sigma_1,\boldsymbol\sigma_1',\boldsymbol\sigma_2} \lambda(\boldsymbol\sigma_2) \psi_1(\boldsymbol\sigma_2;\boldsymbol\sigma_1)\psi_1^*(\boldsymbol\sigma_2;\boldsymbol\sigma_1')
\psi_0(\boldsymbol\sigma_1;\boldsymbol\sigma_0)\psi_0^*(\boldsymbol\sigma_1';\boldsymbol\sigma_0').
\end{align}
To have a clear picture of the calculations, we consider the simple energy values 
\begin{align}
\lambda(\boldsymbol\sigma_2)=-\sum_i h_i\sigma_{2,i}.
\end{align}
Summing over the $\boldsymbol\sigma_2$ variables gives the following transformed energy function:
\begin{align}
\mathsf{H}_1(\boldsymbol\tau_1)\equiv \frac{1}{Z_1}\sum_{\boldsymbol\sigma_2}\lambda(\boldsymbol\sigma_2)e^{\sum_i \tau_{1,i} \sigma_{2,i}}= \frac{1}{\prod_i \cosh B_i^{1,R}} \left(-\sum_i h_i \sinh \tau_{1,i} \prod_{j\neq i} \cosh \tau_{1,i}\right),
\end{align}
with $Z_1=\prod_i (2\cosh B_i^{1,R})$ and $\tau_{1,i}\equiv B_i^{1,R}(\sigma_{1,i}+\sigma_{1,i}')/2-\hat{i}\pi(\sigma_{1,i}-\sigma_{1,i}')/4$.

Again, we see that $\cosh \tau_{1,i} =0$ for $\sigma_{1,i} \neq \sigma_{1,i}$. Thus $\boldsymbol\sigma_1'$ can be different from $\boldsymbol\sigma_1$ at most in one site. As a result, in computing the matrix elements $H_P(\boldsymbol\sigma_0,\boldsymbol\sigma_0')$, we have only two options for the sum over the $\boldsymbol\sigma_1'$: either $\boldsymbol\sigma_1'=\boldsymbol\sigma_1$ or $\boldsymbol\sigma_1'=\boldsymbol\sigma_1^{-i}$ for some $i$. Then, after some simplifications (given in the Appendix), we find
\begin{align}
H_P(\boldsymbol\sigma_0,\boldsymbol\sigma_0')=e^{\hat{i}\Phi(\boldsymbol\sigma_0,\boldsymbol\sigma_0')} \left( \mathsf{H}_0(\boldsymbol\tau_0|\boldsymbol\sigma_1'=\boldsymbol\sigma_1)+\sum_i \mathsf{H}_0(\boldsymbol\tau_0^{-i}|\boldsymbol\sigma_1'=\boldsymbol\sigma_1^{-i}) \right),
\end{align}
where, as before 
\begin{align}
\Phi(\boldsymbol\sigma_0,\boldsymbol\sigma_0')=\Theta^0(\boldsymbol\sigma_0)-\Theta^0(\boldsymbol\sigma_0')+ \sum_i (B_i^{0,I}+\pi/2)\frac{(\sigma_{0,i}-\sigma_{0,i}')}{2}.
\end{align}
We also defined the new conditional functions
\begin{align}
\mathsf{H}_0(\boldsymbol\tau_0|\boldsymbol\sigma_1'=\boldsymbol\sigma_1) &\equiv \frac{1}{Z_0} \sum_{\boldsymbol\sigma_1} \mathsf{H}_1(\boldsymbol\tau_1|\boldsymbol\sigma_1'=\boldsymbol\sigma_1)e^{\sum_i \tau_{0,i}\sigma_{1,i}},\\
\mathsf{H}_0(\boldsymbol\tau_0^{-i}|\boldsymbol\sigma_1'=\boldsymbol\sigma_1^{-i}) &\equiv \frac{1}{Z_0} \sum_{\boldsymbol\sigma_1} \mathsf{H}_1(\boldsymbol\tau_1|\boldsymbol\sigma_1'=\boldsymbol\sigma_1^{-i})e^{\hat{i}\sum_{j\neq i}\Lambda_{ij}^1\sigma_{1,i}\sigma_{1,j}}\times e^{\tau_{0,-i}\sigma_{1,i}+\sum_{j\neq i} \tau_{0,j}\sigma_{1,j}},
\end{align}
with $Z_0=\prod_i (2\cosh B_i^{0,R})$.
These functions depend on $(\boldsymbol\sigma_0,\boldsymbol\sigma_0')$ through the $\boldsymbol\tau_0$ vector, with the following elements:
\begin{align}
\tau_{0,i} &\equiv B_i^{0,R}(\sigma_{0,i}+\sigma_{0,i}')/2-\hat{i}\pi (\sigma_{0,i}-\sigma_{0,i}')/4,\\
\tau_{0,-i} &\equiv B_i^{0,R}(\sigma_{0,i}-\sigma_{0,i}')/2-\hat{i}\pi (\sigma_{0,i}+\sigma_{0,i}')/4+\hat{i}(B_i^{1,I}+\pi/2).
\end{align}

Given the above definitions and energy values, for the simpler case of $\boldsymbol\sigma_1'=\boldsymbol\sigma_1$, we obtain 
\begin{align}
\mathsf{H}_1(\boldsymbol\tau_1|\boldsymbol\sigma_1'=\boldsymbol\sigma_1) &=-\sum_i (h_i \tanh B_i^{1,R})\sigma_{1,i},\\
\mathsf{H}_0(\boldsymbol\tau_0|\boldsymbol\sigma_1'=\boldsymbol\sigma_1) &=-\frac{1}{Z_0}\sum_i h_i \tanh B_i^{1,R}\sinh \tau_{0,i}\prod_{j\neq i}\cosh \tau_{0,j}.
\end{align}
For the case $\boldsymbol\sigma_1'=\boldsymbol\sigma_1^{-i}$, we have 
\begin{align}
\mathsf{H}_1(\boldsymbol\tau_1|\boldsymbol\sigma_1'=\boldsymbol\sigma_1^{-i})=h_i \frac{\hat{i}\sigma_{1,i}}{\cosh B_i^{1,R}}-\sum_{j\neq i}(h_j \tanh B_j^{1,R}) \sigma_{1,j}.
\end{align}
Then, $\mathsf{H}_0(\boldsymbol\tau_0^{-i}|\boldsymbol\sigma_1'=\boldsymbol\sigma_1^{-i})$ is given by the sum of the following expression with the positive and negative signs,
\begin{multline}
\frac{e^{\pm \tau_{0,-i}}}{2\cosh B_i^{0,R}}\Big\{ \frac{\pm \hat{i}h_i}{\cosh B_i^{1,R}}\prod_{j\neq i} \frac{\cosh (\tau_{0,j}\pm \hat{i}\Lambda_{ij}^1)}{\cosh B_j^{0,R}}\\
-\sum_{j\neq i}h_j \tanh B_j^{1,R} \frac{\sinh(\tau_{0,j}\pm \hat{i}\Lambda_{ij}^1)}{\cosh B_j^{0,R}} \prod_{k\neq i,j} \frac{\cosh (\tau_{0,k} \pm \hat{i}\Lambda_{ik}^1)}{\cosh B_k^{0,R}}\Big\}.
\end{multline}
We note that $\cosh \tau_{0,j}=0$ for $\sigma_{0,j}' \neq \sigma_{0,j}$. Therefore, such a spin flip could have a contribution to the Hamiltonian only if $j$ appears in the $\sinh(\tau_{0,j}\pm \hat{i}\Lambda_{ij}^1)$ factors, or $j$ appears in the $\cosh(\tau_{0,j}\pm \hat{i}\Lambda_{ij}^1)$ factors and $\Lambda_{ij}^1$ is nonzero. For each site $i$, the set of possible configurations includes the configuration with no spin flip, the configuration in which only spin $i$ is flipped, the configuration with at least one of the other spins flipped, and so on. In each case, we expand the product factors inside the parentheses to obtain the explicit structure of the interactions. Indeed, accompanied with each $\sigma_j^z$ and $\sigma_j^x$ in the above interaction terms, we have a factor $\tanh B_j^{0,R}$ and $1/\cosh B_j^{0,R}$, respectively. These factors are smaller (in magnitude) than one, and control the degree of locality of the parent Hamiltonian.   

Putting all together, the Hamiltonian that is responsible for the above matrix elements can be written as
\begin{multline}
H_P=-\sum_i h_i^z \sigma_i^z-\sum_i e^{\hat{i}[\sum_{j\neq i}\Lambda_{ij}^0\sigma_i^z\sigma_j^z+B_{i}^{0,I}\sigma_i^z]}h_i^x \sigma_i^x\\
-\sum_i \left( J_i^z\sigma_i^z +\sum_{j\neq i}J_{ij}^{zz} \sigma_i^z\sigma_j^z+ \sum_{j<k: j,k\neq i}J_{ijk}^{zzz} \sigma_i^z\sigma_j^z\sigma_k^z +\cdots \right)\\
-\sum_i e^{\hat{i}[\sum_{j\neq i}\Lambda_{ij}^0\sigma_i^z\sigma_j^z+(B_i^{0,I}+\pi/2)\sigma_i^z]}\left( J_i^{x}\sigma_i^x +\sum_{j\neq i}J_{ij}^{xz} \sigma_i^x\sigma_j^z+ \sum_{j<k: j,k\neq i}J_{ijk}^{xzz}\sigma_i^x\sigma_j^z\sigma_k^z +\cdots \right)-\cdots.
\end{multline}
As mentioned above, the strength of an interaction term like $\prod_{l=1}^m\sigma_{i_l}^z\prod_{l=1}^n\sigma_{j_l}^x$ is proportional to $\prod_{l=1}^m \tanh B_{i_l}^{0,R}/\prod_{l=1}^n\cosh B_{j_l}^{0,R}$, which decreases exponentially with the number of involved spins in the interaction.

\section{Conclusion and Discussion}\label{S5}
In summary, we showed how the energy values that are associated to an ensemble of computationally tractable wave functions affect the structure of the parent Hamiltonian. The findings could have applications in quantum engineering problems where an efficient representation of the Hamiltonian would be very useful in exploring the exponentially large space of the Hamiltonians \cite{MDPZ-aam-2012,WR-qc-2014,ZL-jchem-1995,WZ-chempl-1996}. Consider, for example, the problem of finding an optimal Hamiltonian that maximizes (minimizes) the quantum and thermal expectation of an observable, or maximizes the amplitude of propagating the system from an arbitrary quantum state to another one. The objective function in such problems can be written in terms of the eigenstates and eigenvalues of the Hamiltonian. Here, it is essential to have a concise representation of the (computationally tractable) wave functions and the associated energy values. Furthermore, it would be important to know how the interactions in the Hamiltonian depend on the parameters that specify the energy and wave functions (optimization variables). This knowledge can be used to impose some physical constraints on the optimization variables along with other desirable constraints, for example, on the energy gap of the system \cite{H-prb-2006,H-prb-2007}.

In this paper, we focused on two simple classes of the (multilayer) product states and symmetric tree wave functions.    
The symmetric tree states are more appropriate for describing the spin system in the disordered phase (high temperatures). It would be interesting to investigate the physical consequences of the nonlocal Hamiltonians that we obtained in the study of the tree wave functions. On the other hand, the product states work better than the symmetric tree states in the ordered phase (low temperatures) \cite{R-epl-2013}. The performance of the multilayer product states improves by increasing the number of layers, but, at the same time, the computation time grows exponentially with the number of layers. Finally, we should mention that the approach is not limited to quantum spin systems; a similar study can be done for a system of fermions, for example, with an appropriate set of orthogonal product states \cite{KLC-pra-2013,FK-prb-2011,RZ-prb-2012}. One can indeed start from the (multilayer) Gutzwiller wave functions in the occupation number representation, and obtain a complete set of orthonormal wave functions for fermions. These states then can be used to construct a class of computationally tractable Hamiltonians for fermions on a lattice.

\acknowledgments
We would like to thank V. Karimipour and A. Rezakhani for helpful discussions.

\appendix

\section{Characterizing the parent Hamiltonian: Two-layer product states}\label{app-1}
In this section we give the details of calculations leading to the parent Hamiltonians for the two-layer product states.
We start from the definition of the parent Hamiltonian for a two-layer wave function, 
\begin{align}
H_P(\boldsymbol\sigma_0,\boldsymbol\sigma_0')=\sum_{\boldsymbol\sigma_2,\boldsymbol\sigma_1,\boldsymbol\sigma_1'} \lambda(\boldsymbol\sigma_2) \psi_1(\boldsymbol\sigma_2;\boldsymbol\sigma_1)\psi_1^*(\boldsymbol\sigma_2;\boldsymbol\sigma_1')
\psi_0(\boldsymbol\sigma_1;\boldsymbol\sigma_0)\psi_0^*(\boldsymbol\sigma_1';\boldsymbol\sigma_0').
\end{align}
For a two-layer wave function of product states, we have
\begin{multline}
\psi_0(\boldsymbol\sigma_1;\boldsymbol\sigma_0)\psi_0^*(\boldsymbol\sigma_1';\boldsymbol\sigma_0')=\frac{e^{\hat{i}[\Theta^0(\boldsymbol\sigma_0)-\Theta^0(\boldsymbol\sigma_0')]}}{\prod_i (2\cosh B_i^{0,R})}
\times \exp\left(\sum_i B_i^{0,R}(\sigma_{0,i}\sigma_{1,i}+\sigma_{0,i}'\sigma_{1,i}')/2\right)
\\ \times \exp\left(\sum_i\hat{i}B_i^{0,I}(\sigma_{0,i}-\sigma_{0,i}')/2\right) \times \exp\left(\hat{i}\pi\sum_i[\sigma_{0,i}(1-\sigma_{1,i})-\sigma_{0,i}'(1-\sigma_{1,i}')]/4\right),
\end{multline}
and similarly for $\psi_1(\boldsymbol\sigma_2;\boldsymbol\sigma_1)\psi_1^*(\boldsymbol\sigma_2';\boldsymbol\sigma_1')$.
Then, the matrix elements of the parent Hamiltonian are written as
\begin{multline}
H_P(\boldsymbol\sigma_0,\boldsymbol\sigma_0')=\frac{e^{\hat{i}[\Theta^0(\boldsymbol\sigma_0)-\Theta^0(\boldsymbol\sigma_0')]}}{Z_0Z_1}\sum_{\boldsymbol\sigma_2,\boldsymbol\sigma_1,\boldsymbol\sigma_1'} 
\lambda(\boldsymbol\sigma_2)e^{\hat{i}[\Theta^1(\boldsymbol\sigma_1)-\Theta^1(\boldsymbol\sigma_1')]}\\ \times
\exp\left(\sum_i B_i^{1,R}(\sigma_{1,i}\sigma_{2,i}+\sigma_{1,i}'\sigma_{2,i})/2\right)\times
\exp\left(\sum_i B_i^{0,R}(\sigma_{0,i}\sigma_{1,i}+\sigma_{0,i}'\sigma_{1,i}')/2\right)
\\ \times \exp\left(\hat{i}\sum_i B_i^{1,I}(\sigma_{1i}-\sigma_{1,i}')/2\right) \times \exp\left(\hat{i}\pi\sum_i(\sigma_{1,i}-\sigma_{1,i}')(1-\sigma_{2,i})/4\right) 
\\ \times \exp\left(\hat{i}\sum_i B_i^{0,I}(\sigma_{0,i}-\sigma_{0,i}')/2 \right)\times \exp\left(\hat{i}\pi\sum_i[\sigma_{0,i}(1-\sigma_{1,i})-\sigma_{0,i}'(1-\sigma_{1,i}')]/4 \right),
\end{multline}
where $Z_0=\prod_i (2\cosh B_i^{0,R})$ and $Z_1=\prod_i (2\cosh B_i^{1,R})$.
Let us define $\tau_{1,i}\equiv B_i^{1,R}(\sigma_{1,i}+\sigma_{1,i}')/2-\hat{i}\pi(\sigma_{1,i}-\sigma_{1,i}')/4$, and
the transformed energy function
\begin{align}
\mathsf{H}_1(\boldsymbol\tau_1)\equiv \frac{1}{Z_1}\sum_{\boldsymbol\sigma_2}\lambda(\boldsymbol\sigma_2)e^{\sum_i \tau_{1,i} \sigma_{2,i}}.
\end{align}
Using the above definitions, we get
\begin{multline}
H_P(\boldsymbol\sigma_0,\boldsymbol\sigma_0')=\frac{e^{\hat{i}[\Theta^0(\boldsymbol\sigma_0)-\Theta^0(\boldsymbol\sigma_0')]}}{Z_0}\sum_{\boldsymbol\sigma_1,\boldsymbol\sigma_1'} 
\mathsf{H}_1(\boldsymbol\tau_1)e^{\hat{i}[\Theta^1(\boldsymbol\sigma_1)-\Theta^1(\boldsymbol\sigma_1')]}
\\ \times \exp\left(\sum_i B_i^{0,R}(\sigma_{0,i}\sigma_{1,i}+\sigma_{0,i}'\sigma_{1,i}')/2\right)
\\ \times 
\exp\left(\hat{i}\sum_iB_i^{1,I}(\sigma_{1,i}-\sigma_{1,i}')/2\right) \times \exp\left(\hat{i}\pi\sum_i(\sigma_{1,i}-\sigma_{1,i}')/4\right)
\\ \times \exp\left(\hat{i}\sum_i B_i^{0,I}(\sigma_{0,i}-\sigma_{0,i}')/2 \right)\times \exp\left(\hat{i}\pi\sum_i[\sigma_{0,i}(1-\sigma_{1,i})-\sigma_{0,i}'(1-\sigma_{1,i}')]/4 \right).
\end{multline}

In the following, we consider the simple energy values 
\begin{align}
\lambda(\boldsymbol\sigma_2)=-\sum_i h_i\sigma_{2,i},
\end{align}
which give the transformed energy function
\begin{align}
\mathsf{H}_1(\boldsymbol\tau_1)= \frac{1}{\prod_i \cosh B_i^{1,R}} \left(-\sum_i h_i \sinh \tau_{1,i} \prod_{j\neq i} \cosh \tau_{1,i}\right).
\end{align}
Recall that $\boldsymbol\tau_1$ is a function of $(\boldsymbol\sigma_1,\boldsymbol\sigma_1')$, and that $\cosh \tau_{1,i} =0$ for $\sigma_{1,i}' \neq \sigma_{1,i}$. Thus, $\boldsymbol\sigma_1'$ can be different from $\boldsymbol\sigma_1$ at most in one site. As a result, in computing the matrix elements $H_P(\boldsymbol\sigma_0,\boldsymbol\sigma_0')$, we have only two options for the sum over the $\boldsymbol\sigma_1'$; either $\boldsymbol\sigma_1'=\boldsymbol\sigma_1$ or $\boldsymbol\sigma_1'=\boldsymbol\sigma_1^{-i}$ for some spin $i$. It means that we can write
\begin{align}
H_P(\boldsymbol\sigma_0,\boldsymbol\sigma_0')=e^{\hat{i}\Phi(\boldsymbol\sigma_0,\boldsymbol\sigma_0')} \left( \mathsf{H}_0(\boldsymbol\tau_0|\boldsymbol\sigma_1'=\boldsymbol\sigma_1)+\sum_i \mathsf{H}_0(\boldsymbol\tau_0^{-i}|\boldsymbol\sigma_1'=\boldsymbol\sigma_1^{-i}) \right),
\end{align}
where, as before 
\begin{align}
\Phi(\boldsymbol\sigma_0,\boldsymbol\sigma_0')=\Theta^0(\boldsymbol\sigma_0)-\Theta^0(\boldsymbol\sigma_0')+ \sum_i (B_i^{0,I}+\pi/2)\frac{(\sigma_{0,i}-\sigma_{0,i}')}{2},
\end{align}
and we defined the conditional energy functions 
\begin{align}
\mathsf{H}_0(\boldsymbol\tau_0|\boldsymbol\sigma_1'=\boldsymbol\sigma_1) &\equiv \frac{1}{Z_0} \sum_{\boldsymbol\sigma_1} \mathsf{H}_1(\boldsymbol\tau_1|\boldsymbol\sigma_1'=\boldsymbol\sigma_1)e^{\sum_i \tau_{0,i}\sigma_{1,i}},\\
\mathsf{H}_0(\boldsymbol\tau_0^{-i}|\boldsymbol\sigma_1'=\boldsymbol\sigma_1^{-i}) &\equiv \frac{1}{Z_0} \sum_{\boldsymbol\sigma_1} \mathsf{H}_1(\boldsymbol\tau_1|\boldsymbol\sigma_1'=\boldsymbol\sigma_1^{-i})e^{\hat{i}\sum_{j\neq i}\Lambda_{ij}^1\sigma_{1,i}\sigma_{1,j}}\times e^{\tau_{0,-i}\sigma_{1,i}+\sum_{j\neq i} \tau_{0,j}\sigma_{1,j}}.
\end{align}
These functions depend on $(\boldsymbol\sigma_0,\boldsymbol\sigma_0')$ through the $\boldsymbol\tau_0$ vector, with the following elements
\begin{align}
\tau_{0,i} &\equiv B_i^{0,R}(\sigma_{0,i}+\sigma_{0,i}')/2-\hat{i}\pi (\sigma_{0,i}-\sigma_{0,i}')/4,\\
\tau_{0,-i} &\equiv B_i^{0,R}(\sigma_{0,i}-\sigma_{0,i}')/2-\hat{i}\pi (\sigma_{0,i}+\sigma_{0,i}')/4+\hat{i}(B_i^{1,I}+\pi/2).
\end{align}

Given the above definitions and energy values, we obtain
\begin{align}
\mathsf{H}_1(\boldsymbol\tau_1|\boldsymbol\sigma_1'=\boldsymbol\sigma_1)=-\sum_i (h_i \tanh B_i^{1,R}) \sigma_{1,i},
\end{align}
and then 
\begin{multline}
\mathsf{H}_0(\boldsymbol\tau_0|\boldsymbol\sigma_1'=\boldsymbol\sigma_1) =-\frac{1}{Z_0}\sum_i h_i \tanh B_i^{1,R} \sinh \tau_{0,i}\prod_{j\neq i}\cosh \tau_{0,j}\\ 
= -\sum_i h_i \tanh B_i^{1,R} \left( \tanh B_i^{0,R}\delta_{\boldsymbol\sigma_0',\boldsymbol\sigma_0}-\frac{\hat{i} \delta_{\boldsymbol\sigma_0',\boldsymbol\sigma_0^{-i}}}{\cosh B_i^{0,R}} \right)\sigma_{0,i}.
\end{multline}
The Hamiltonian that is responsible for this part can be written as
\begin{align}
H_P=-\sum_i h_i^z \sigma_i^z-\sum_i e^{\hat{i}[\sum_{j\neq i}\Lambda_{ij}^0\sigma_i^z\sigma_j^z+B_{i}^{0,I}\sigma_i^z]}h_i^x \sigma_i^x,
\end{align}
with $h_i^z \equiv h_i\tanh B_i^{1,R}\tanh B_i^{0,R}$ and $h_i^x \equiv h_i\tanh B_i^{1,R}/\cosh B_i^{0,R}$.

For the case $\boldsymbol\sigma_1'=\boldsymbol\sigma_1^{-i}$, we have 
\begin{align}
\mathsf{H}_1(\boldsymbol\tau_1|\boldsymbol\sigma_1'=\boldsymbol\sigma_1^{-i})=-h_i \frac{-\hat{i}\sigma_{1,i}}{\cosh B_i^{1,R}}-\sum_{j\neq i}h_j \tanh B_j^{1,R} \sigma_{1,j},
\end{align}
and $\mathsf{H}_0(\boldsymbol\tau_0^{-i}|\boldsymbol\sigma_1'=\boldsymbol\sigma_1^{-i})$ is given by
\begin{multline}
\frac{e^{\tau_{0,-i}}}{2\cosh B_i^{0,R}}\Big\{ \frac{\hat{i}h_i}{\cosh B_i^{1,R}}\prod_{j\neq i} \frac{\cosh (\tau_{0,j}+\hat{i}\Lambda_{ij}^1)}{\cosh B_j^{0,R}}\\
-\sum_{j\neq i}h_j \tanh B_j^{1,R} \frac{\sinh(\tau_{0,j}+\hat{i}\Lambda_{ij}^1)}{\cosh B_j^{0,R}} \prod_{k\neq i,j} \frac{\cosh (\tau_{0,k}+\hat{i}\Lambda_{ik}^1)}{\cosh B_k^{0,R}}\Big\},
\end{multline}
plus
\begin{multline}
\frac{e^{-\tau_{0,-i}}}{2\cosh B_i^{0,R}}\Big\{ \frac{-\hat{i}h_i}{\cosh B_i^{1,R}}\prod_{j\neq i} \frac{\cosh (\tau_{0,j}-\hat{i}\Lambda_{ij}^1)}{\cosh B_j^{0,R}}\\
-\sum_{j\neq i}h_j \tanh B_j^{1,R} \frac{\sinh(\tau_{0,j}-\hat{i}\Lambda_{ij}^1)}{\cosh B_j^{0,R}} \prod_{k\neq i,j} \frac{\cosh (\tau_{0,k}-\hat{i}\Lambda_{ik}^1)}{\cosh B_k^{0,R}}\Big\}.
\end{multline}
We note that $\cosh \tau_{0,j}=0$ for $\sigma_{0,j}' \neq \sigma_{0,j}$. Therefore, such a spin flip could have a contribution to the Hamiltonian only if $\Lambda_{ij}^1$ is nonzero. Moreover, the factors appearing in the above expression can be rewritten as  
\begin{align}
\frac{\cosh (\tau_{0,j}\pm \hat{i}\Lambda_{ij}^1)}{\cosh B_j^{0,R}}=
\left\{
       \begin{array}{ll}
       \cos \Lambda_{ij}^1\pm \hat{i}\sigma_{0,j}\tanh B_j^{0,R}\sin \Lambda_{ij}^1, & \hbox{if $\sigma_{0,j}'=\sigma_{0,j}$;} \\
        \pm \frac{\sigma_{0,j}}{\cosh B_j^{0,R}}\sin \Lambda_{ij}^1, & \hbox{otherwise.}
       \end{array}
       \right.
\end{align}
and
\begin{align}
\frac{\sinh (\tau_{0,j}\pm \hat{i}\Lambda_{ij}^1)}{\cosh B_j^{0,R}}=
\left\{
       \begin{array}{ll}
       \sigma_{0,j} \tanh B_j^{0,R}\cos \Lambda_{ij}^1 \pm \hat{i}\sin \Lambda_{ij}^1, & \hbox{if $\sigma_{0,j}'=\sigma_{0,j}$;} \\
       -\frac{\hat{i}\sigma_{0,j}}{\cosh B_j^{0,R}}\cos \Lambda_{ij}^1, & \hbox{otherwise.}
       \end{array}
       \right.
\end{align}
To write the parent Hamiltonian, we need to consider all the contributions from site $i$. This includes the configuration with no spin flip, the configuration in which only spin $i$ is flipped, the configuration in which at least one of the other spins is flipped, and so on. In each case, we expand the product factors to obtain the explicit structure of the interactions.

Putting all together, the Hamiltonian that is responsible for the above matrix elements can be written as
\begin{multline}
H_P=-\sum_i \left( J_i^z\sigma_i^z +\sum_{j\neq i}J_{ij}^{zz} \sigma_i^z\sigma_j^z+ \sum_{j<k: j,k\neq i}J_{ijk}^{zzz} \sigma_i^z\sigma_j^z\sigma_k^z +\cdots \right)\\
-\sum_i e^{\hat{i}[\sum_{j\neq i}\Lambda_{ij}^0\sigma_i^z\sigma_j^z+(B_i^{0,I}+\pi/2)\sigma_i^z]}\left( J_i^{x}\sigma_i^x +\sum_{j\neq i}J_{ij}^{xz} \sigma_i^x\sigma_j^z+ \sum_{j<k: j,k\neq i}J_{ijk}^{xzz}\sigma_i^x\sigma_j^z\sigma_k^z +\cdots \right)-\cdots.
\end{multline}

\end{document}